# TOWARDS ROBUST MISPRONUNCIATION DETECTION AND DIAGNOSIS FOR L2 ENGLISH LEARNERS WITH ACCENT-MODULATING METHODS

*Shao-Wei Fan Jiang, Bi-Cheng Yan, Tien-Hong Lo, Fu-An Chao, Berlin Chen*

National Taiwan Normal University
{swfanjiang, bicheng, teinhonglo, fuann, berlin}@ntnu.edu.tw

**ABSTRACT**

With the acceleration of globalization, more and more people are willing or required to learn second languages (L2). One of the major remaining challenges facing current mispronunciation and diagnosis (MDD) models for use in computer-assisted pronunciation training (CAPT) is to handle speech from L2 learners with a diverse set of accents. In this paper, we set out to mitigate the adverse effects of accent variety in building an L2 English MDD system with end-to-end (E2E) neural models. To this end, we first propose an effective modeling framework that infuses accent features into an E2E MDD model, thereby making the model more accent-aware. Going a step further, we design and present disparate accent-aware modules to perform accent-aware modulation of acoustic features in a finer-grained manner, so as to enhance the discriminating capability of the resulting MDD model. Extensive sets of experiments conducted on the L2-ARCTIC benchmark dataset show the merits of our MDD model, in comparison to some existing E2E-based strong baselines and the celebrated pronunciation scoring based method.

***Index Terms:*** mispronunciation detection and diagnosis, accented speech, computer-assisted pronunciation training, accent modeling, multi-task learning

## 1. INTRODUCTION

The tide of globalization has underscored the importance of foreign language proficiency. Development of computer-assisted pronunciation training (CAPT) systems opens up new possibilities to enhance L2 pronunciation skills in an effective and stress-free manner. Furthermore, more and more standardized examinations also use CAPT systems to assist L2 proficiency evaluations (e.g., TOEFL [1] and AZELLA [2]).

As an integral component of a CAPT system, mispronunciation detection and diagnosis (MDD) manages to detect phone- or word-level pronunciation errors from the speech of L2 learners, and in turn provide them with the associated diagnosis feedback [3]. For this purpose, several research directions towards effective MDD have been explored in the literature. One classic category of mispronunciation detection methods is pronunciation scoring, which computes phone-level pronunciation scores based on confidence measures derived from hybrid deep neural network-hidden Markov model (DNN-HMM) based automatic speech recognition (ASR), including phone durations, phone posterior probability scores and segment duration scores, and among others [4], [5]. Goodness of pronunciation (GOP) and its variants are the most representative methods of this category [6], [7]. However, this category of methods cannot readily provide L2 learners with tangible mispronunciation diagnosis. Another prevalent category of methods seeks to assess the details of mispronunciations in order to provide L2 learners with diagnosis feedback about some specific errors such as phone substitutions, deletions and insertions [8]. A prominent instantiation of this category is the extended recognition network (ERN) method [9], [10]. ERN augments the decoding network of ASR with phonological rules. By comparison between an ASR output and the corresponding text prompt, ERN is amenable to offering appropriate diagnosis feedback of mispronunciations. However, it is practically difficult to enumerate and include sufficient phonological rules into the decoding network for different L1-L2 language pairs. Yet, a more recent trend is to make the end-to-end (E2E) neural ASR paradigm straightforwardly applicable to MDD, which in essence employs a free-phone recognition topology implemented with deep neural networks like connectionist temporal classification (CTC), attention-based model or their hybrids (denoted by hybrid CTT-ATT) [11], [12], [13]. Methods stemming from this line of research have shown promising mispronunciation detection performance, in relation to the GOP-based method that builds on the DNN-HMM based ASR paradigm.

On a separate front, as is well known, L2 English learners may have a wide variety of accents, which is usually influenced by their mother-tongue languages. However, current E2E approaches to MDD largely deal with multiple accents by simply pooling correctly-pronounced utterances from several accents (including those of L1 English speakers) to train a single, one-size-fits-all MDD model [15], [16], [17], which might inevitably weaken the discriminating power of the resulting model. In view of the above observations, we embark on an effort to build an MDD system with E2E neural models for English L2 learners [18], [19], [20] which is more resilient to the undesirable influence of accent variety. In so doing, we first propose an effective modeling framework that integrates accent features into the MDD model to make the model more accent-aware. Going a step further, we design and implement different accent aware modules that perform accent-aware modulation of acoustic features, viz. the hidden acoustic embeddings of an input utterance, in a finer-grained manner to enhance the discriminating capability of the MDD model.

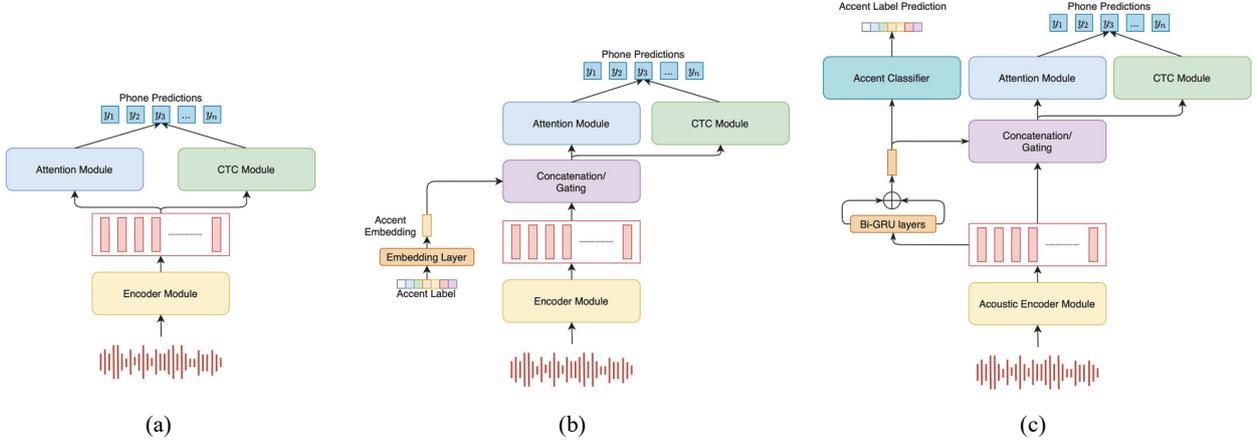

Figure 1: *Different E2E MDD models for mispronunciation detection: (a) the baseline CTC-ATT model; (b) the specific accent-aware model; (c) the multi-task accent-aware model.*

The rest of the paper is organized as follows. In Section 2, we shed light on the formulations and architectures of different baseline E2E MDD models. After that, we will introduce our proposed accent-aware MDD models in Section 3, followed by the experimental setup and results presented in Section 4. Finally, we conclude the paper and suggest avenues for future work in Section 5.

## 2. HYBRID CTC-ATT MODEL FOR MDD

A typical hybrid CTC-ATT MDD model consists of four modules, as depicted in Figure 1(a): 1) an encoder module that is shared across CTC and the attention-based model. The encoder module extracts intermediate representation $H = (\mathbf{h}_1, \ldots, \mathbf{h}_S)$ from an input acoustic feature vector sequence $O = (\mathbf{o}_1, \ldots, \mathbf{o}_T)$ (e.g., log Mel-filter-bank feature vectors) through a stack of, for example, convolution or recurrent networks, where $S \leq T$ due to the subsampling operation; 2) an attention module that calculates a fixed-length context vector $\mathbf{c}_l$ by summarizing the output of the encoder module at each output step for $l \in [1, \ldots, L]$, finding out relevant parts of the encoder state sequence to be attended for predicting an output phone symbol $y_l$, where the output symbol sequence $Y = (y_1, \ldots, y_L)$ belongs to a canonical phone set $\mathcal{U}$; 3) Given the context vector $\mathbf{c}_l$ and the history of partial diagnostic results $y_{1:l-1}$, a decoder module updates its hidden state $\mathbf{q}_l$ autoregressively and estimates the next phone symbol $y_l$; 4) The CTC module that predicts a frame-level alignment between the hidden vector sequence $\mathbf{h}_{1:S}^E$ and the output sequence Y by additionally introducing a special <blank> token. It can reduce irregular alignments substantially during the training and test phases.

The training objective function of the hybrid CTC-ATT model is to maximize a logarithmic linear combination of the posterior probabilities predicted by CTC and the attention-based model, i.e., $P_{CTC}(Y|O)$ and $P_{att}(Y|O)$:

$$\mathcal{L}_{\text{ctc-att}} = \alpha \log P_{\text{ctc}}(Y|O) + (1-\alpha) \log P_{\text{att}}(Y|O), \quad (1)$$

$$P_{\text{ctc}}(Y|O) = \sum_Z P(Y|Z,O)P(Z|O)$$
$$\approx \sum_Z P(Y|Z)P(Z|O), \quad (2)$$

$$P_{att}(Y|O) = \prod_{l=1}^{L} P(y_l|y_{1:l-1}, O). \quad (3)$$

where the frame-wise latent variable sequences Z belongs to a canonical phone set $\mathcal{U}$ augmented with and the additional <blank> label, which facilitates CTC to enforce a monotonic behavior of phone-level alignments. Eq. (2) is a neat formulation of the CTC model, which is derived from the assumption that the symbol translation model $P(Y|Z)$ is conditionally independent of the input sequence O [22]. The linear combination weight $\alpha$ in Eq. (1) is a hyperparameter used to linearly interpolate the two posterior probabilities. In our experiments, the default value of $\alpha$ is set equal to 0.3.

# 3. ACCENT-MODULATED ENCODERS FOR MDD

Since there is mismatch of the acoustic traits between native speakers and L2 speakers with different mother-tongue accents, the E2E-based MDD model simply trained on a mixture of a large quantity of native utterances and a small quantity of L2 utterances of different accents might not perform well in realistic scenarios. A natural way to address the above issue is to encourage the encoder module of the E2E-based MDD model to extract the intermediate representation of a learner's speech that carries accent-aware acoustic features, which are in turn expected to facilitate the decoder to precisely determine the correctness of the content that is uttered by the learner in response to the text prompt. To this end, we design two different accent-aware encoder modules to better account for the acoustic variety of L2 speakers with different mother-tongue accents.

## 3.1 Accented-aware modeling

In the first proposed approach, we assume that the information about the accent of an utterance is known a priori in both the training and test phases. As can be seen from Figure 1 (b), the embedding vector of the accent label of an input utterance is fed into the encoder module to encourage it generate accent-modulated intermediate acoustic vectors. To this end, two disparate embodiments of this approach are developed to fuse together accent information and acoustic features, namely the accent-modulated concatenation (AMC) mechanism and accent-modulated gating (AMG) mechanism.

For the AMC mechanism, as can be seen from Figure 2 (a), we simply concatenate the utterance-level embedding vector $\mathbf{a}_E$ with each intermediate acoustic vector $\mathbf{h}_S$ to form a new intermediate acoustic vector $\mathbf{h}'_S$ that incorporates the sentence-level accent information:

$$\mathbf{a}_E = \text{Embedding}(\mathbf{a}_{one-hot}) \quad (4)$$

$$\mathbf{h}'_S = \text{concat}(\mathbf{h}_S, \mathbf{a}_E)W_1 + \mathbf{b}_1 \quad (5)$$

where $\mathbf{a}_{one-hot}$ and $\mathbf{a}_E$ are the one-hot representation and distributed representation (embedding) of the accent label of the input utterance, respectively; and $W_1$ and $\mathbf{b}_1$ are transformation matrix and bias vector for dimensionality reduction, respectively.

Building on top of the AMC mechanism, as can be seen from Figure 2 (b), we design and develop a gating mechanism (i.e., AMG) alternatively to perform finer-grained accent-aware modulation of the intermediate acoustic vector, with the hope to further enhance the discriminating capability of the MDD model. The formulations of AMG are concisely expressed as follows:

$$\mathbf{g}_S = \text{Sigmoid}([\mathbf{h}_S; \mathbf{a}_E]W_2 + \mathbf{b}_2) \quad (6)$$

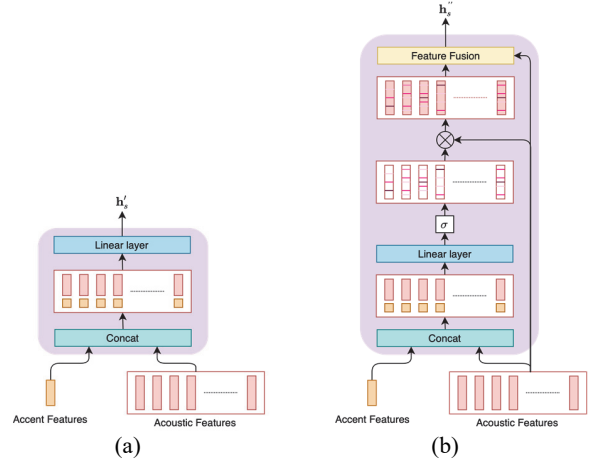

Figure 2: *Variants of accent-modulating: (a) accent-modulating concatenation (AMC) mechanism; (b) accent-modulating gate (AMG) mechanism*

$$\mathbf{h}''_S = \text{Relu}(\mathbf{h}_S + [\mathbf{h}_S \odot \mathbf{g}_S]W_3 + \mathbf{b}_3) \quad (7)$$

where $\odot$ represents the elementwise dot production and $W_2$, $W_3$, $\mathbf{b}_2$ and $\mathbf{b}_3$ are trainable parameters. The functionality of Eq. (6) is to model the correlation and complementarity between an intermediate acoustic vector $\mathbf{h}_S$ and the utterance-level accent vector $\mathbf{a}_E$ at timestep $s$, producing a gating vector $\mathbf{g}_S$ with the sigmoid function. After that, Eq. (7) seeks to obtain a modulated intermediate acoustic vector $\mathbf{h}_S^{E'}$ with the gating vector $\mathbf{g}_S$ and the Relu function.

## 3.2 Soft accent-aware modeling

Since the accent information about an L2 learner is not always readily available in the test phase under realistic scenarios, we instead to design and build a soft accent-aware encoder module on the basis of multi-task training. More specifically, the multi-task training process jointly optimizes the prediction objectives of the corresponding canonical phone labels (the primary task) and accent labels (the auxiliary task) of all training utterances, as depicted in Figure 1 (c). Note here that the extra accent information fused into the encoder module when using AMC or AMG is the output of the last hidden layer of the accent-classification auxiliary task, which can be thought of as a kind of soft accent-aware modeling. This way, we only require to know the accent category of the training utterances when training the MDD model, while the test utterances being kept accent-agnostic.

Finally, a bit of terminology: the secondary task employs an accent classifier whose output layer (with the softmax activation) estimates the accent label of an utterance, and the learned gradients of the accent classifier encourages the shared intermediate acoustic vectors $\mathbf{h}_{1:S}$ to carry acoustic traits that are more accent-aware. In implementation, the accent classifier summarizes the intermediate acoustic vectors $\mathbf{h}_{1:S}$ of an utterance by a stack of multiple-layer bidirectional gated recurrent unit (GRU) neural network to produce an accent embedding $\mathbf{a}'_E$, which is in turn used to predict an utterance-level accent label. The joint training objective function of the MDD model and the accent classifier is expressed by:

$$\mathcal{L} = (1-\beta) \cdot \mathcal{L}_{\text{CTC-ATT}} + \beta \cdot \mathcal{L}_{\text{AC}}. \quad (8)$$

where $\mathcal{L}_{\text{CTC-ATT}}$ is the loss function of the hybrid CTC-ATT model for dictating canonical phones, and $\mathcal{L}_{\text{AC}}$ is the loss function for the accent classification. The hyperparameter $\beta > 0$ is the interpolation weight that controls how much we rely on the gradients backpropagated from the accent classification task. Thereafter, we denote the AMC and AMG mechanisms with such a soft accent-aware modeling strategy by AMC-S and AMG-S, respectively.

Table 1: *Statistics of the experimental speech corpora.*

|    | Corpus   | subsets | Spks. | Utters. | Hrs. |
|----|----------|---------|-------|---------|------|
| L1 | TIMIT    | Train   | 462   | 3696    | 3.15 |
|    |          | Dev.    | 50    | 400     | 0.34 |
|    |          | Test    | 24    | 192     | 0.16 |
| L2 | L2-ARCTIC | Train  | 17    | 2549    | 2.66 |
|    |          | Dev.    | 1     | 150     | 0.12 |
|    |          | Test    | 6     | 900     | 0.88 |

Table 2: *The confusion matrix of mispronunciation detection and diagnosis task.*

| Total Conditions | | Ground Truth | |
|---|---|---|---|
| | | CP | MP |
| Model Prediction | CP | True Positive (TP) | False Positive (FP) |
| | MP | False Negative (FN) | True Negative (TN) |

## 4. EXPERIMENTS

### 4.1 Speech corpora and experimental setup

We conduct our mispronunciation detection experiments on the L2-ARCTIC dataset [23]. The L2-ARCTIC dataset is an open-access L2-English speech corpus compiled for research in CAPT, accent conversion, and others. L2-ARCTIC contains correctly pronounced utterances and mispronounced utterances of 24 non-native speakers (12 males and 12 females), whose mother-tongue languages include Hindi, Korean, Mandarin, Spanish, Arabic and Vietnamese. A suitable amount of native (L1) English speech data compiled from the TIMIT corpus [24] (composed of 630 speakers) was used to bootstrap the training of the various E2E MDD models. To unify the phone sets of these two corpora, we first followed the definition of the CMU pronunciation dictionary to obtain an inventory of 48 canonical phones and in turn converted them into a condensed set of 39 canonical phones.

Next, we divided these two corpora into training, development and test sets, respectively; in particular, the setting of the mispronunciation detection experiments on L2-ARCTIC followed the recipe provided by [12]. Table 1 summarizes detailed statistics of these speech datasets.

The encoder modules of our various E2E MDD models were all composed of a stack of Transformer-based components (each of which is constructed using eight-head attention), 2,048 linear-layer hidden units and 512 output units. Meanwhile, the input to encoder module were 80-dimensional log Mel-filter-bank feature vectors which were extracted every 10 ms with a Hanning window size of 25 ms [25]. Next, the decoder module of our various E2E MDD models all consisted of Transformer-based components (with 8 eight-head attention) and 2,048 linear-layer hidden units. Furthermore, the accent classifier was trained on the training set of L2-ARCTIC that includes 7 accents, namely, English, Hindi, Korean, Mandarin, Spanish, Arabic and Vietnamese.

For AMC and AMG, the dimensionality of the accent vector is set to 128 dimensions. More specifically, the accent classifier involved in AMC-S and AMG-S was built with a multi-layer bidirectional gated recurrent unit (Bi-GRU) neural network with 7 hidden units followed by a softmax output layer, from which the 128-dimensional output of the last hidden layer (prior to the output layer) was taken as the soft accent-aware information for use in the AMC-S and AMG-S mechanisms, respectively.

### 4.2 Performance evaluation metrics

For the mispronunciation detection experiments, we follow the hierarchical evaluation structure adopted in [3], while the corresponding confusion matrix for four test conditions is illustrated in Table 2, where the CP and MP indicate correct pronunciation and mispronunciation conditions, respectively. Based on the statistics accumulated from the four test conditions, we can calculate the values of different metrics like recall (RE; TN/(FP+TN)), precision (PR; TN/(FN+TN)) and the F-1 measure (F-1; the harmonic mean of the precision and recall), so as to evaluate the performance of mispronunciation detection.

Furthermore, to calculate the diagnosis accuracy rate (DAR), we focus on the cases of TN and consider it as combination of the numbers of correct diagnosis (CD) and incorrect diagnosis (ID). The formula for calculating DAR of mispronunciations is defined by:

$$\text{DAR} = \frac{\text{CD}}{\text{CD} + \text{ID}}. \quad (9)$$

### 4.3 Experimental results

In our first set of experiments, we evaluate the performance levels of our accent-aware E2E models for mispronunciation

Table 3: *Performance on mispronunciation detection for our models, viz. AMC and AMG, in comparison to CTC-ATT.*

| Models  | RE (%) | PR (%) | F1 (%) |
|---------|--------|--------|--------|
| CTC-ATT | 47.56  | 46.25  | 46.95  |
| AMC     | 51.32  | 48.96  | 50.11  |
| AMG     | 49.85  | 49.47  | 49.66  |

Table 4: *Performance on mispronunciation detection for AMC-S and AMG-S as a function of the number of Bi-GRU layers employed by the accent classifier.*

| Models | # of layers | RE (%) | PR (%) | F1 (%) |
|--------|-------------|--------|--------|--------|
| AMC-S  | 1           | 49.34  | 47.96  | 48.64  |
|        | 2           | 50.39  | 48.66  | 49.51  |
|        | 3           | 49.32  | 48.39  | 48.85  |
| AMG-S  | 1           | 49.90  | 47.17  | 48.49  |
|        | 2           | 48.39  | 47.46  | 47.92  |
|        | 3           | 50.68  | 49.57  | 50.12  |

Table 5: *Performance comparison of our models with several existing models for mispronunciation detection.*

| Models       | RE (%) | PR (%) | F1 (%) |
|--------------|--------|--------|--------|
| GOP          | 50.15  | 46.99  | 48.52  |
| CNN-RNN-CTC* | 67.29  | 34.88  | 45.94  |
| CTC-ATT      | 47.56  | 46.25  | 46.95  |
| AMC          | 51.32  | 48.96  | 50.11  |
| AMG          | 49.85  | 49.47  | 49.66  |
| AMC-S        | 50.39  | 48.66  | 49.51  |
| AMG-S        | 50.68  | 49.57  | 50.12  |

Note: *We reproduced the E2E phone recognition framework reported in [11].

detection, viz. AMC and AMG, in relation to the baseline hybrid CTC-ATT based model (denoted by CTC-ATT). As can be seen from Table 3, both AMC and AMG can achieve substantial performance improvements over CTC-ATT in terms of the recall, precision and F-1 evaluation metrics. Furthermore, AMC and AMG appear to perform on par with each other, while the latter additionally invokes a more sophisticated gating mechanism to fuse together the intermediate acoustic vectors and accent embedding.

After that, we turn to evaluate the performance of different numbers of Bi-GRU layers employed by the accent classifier for the soft accent-aware mispronunciation detection models, viz. AMC-S and AMG-S. Inspection of Table 4 reveals two noteworthy points. First, AMC-S and AMG-S achieve their best results when the number of the Bi-GRU layers is set to 2 and 3, respectively. Second, AMG-S gives superior performance over AMC-S when test utterances are simply assumed to be accent-agnostic. AMG-S also keeps abreast with AMG and AMC, while the latter two methods require that the accent information of a test utterance is known a priori.

We then assess the efficacy of our methods on mispronunciation detection in relation to the well-studied GOP-based model [23] and the CNN-RNN-CTC neural model proposed in [11], whose results are shown in Table 5. As can be seen, our proposed accent-aware E2E neural models all surpass the GOP-based model and the CNN-RNN-CTC model, confirming the promise of using accent-aware cues to improve the performance of mispronunciation detection. Furthermore, AMG-S that leverages a gating mechanism to integrate extracted accent-aware information into CTC-ATT stands out in performance, as well as shows its practicality.

In the last set of experiments, we evaluate the mispronunciation diagnosis performance of different E2E neural models. The corresponding diagnosis accuracy rates (DAR) for CNN-RNN-CTC, CTC-ATT, AMC, AMG, AMC-S and AMG-S model are 57.59%, 70.74%, 73.31%, 72.28%, 72.86%, 73.75%, respectively. Though the DAR results for these neural models seem to be still far from satisfactory, AMG-S perform the best among these models.

## 5. CONCLUSION AND FUTURE WORK

In this paper we have designed and developed variants of E2E neural methods for mispronunciation detection and diagnosis (MDD) which attempt to mitigate the impacts of accent variety on the MDD task of a CAPT system. The experimental results on the L2-ARCTIC benchmark dataset show that our methods can offer remarked performance improvements over the baseline E2E model and the well-established GOP method. As to future work, we plan to combine the strengths of our models and others state-of-the-art models [26], [27], [28], [29], as well as to enhance the model robustness for practical applications [30].

## 5. ACKNOWLEDGEMENT

This research is supported in part by Chunghwa Telecom Laboratories under Grant Number TL-110-D301, and by the National Science Council, Taiwan under Grant Number MOST 109-2634-F-008-006- through Pervasive Artificial Intelligence Research (PAIR) Labs, Taiwan, and Grant Numbers MOST 108-2221-E-003-005-MY3 and MOST 109-2221-E-003-020-MY3. Any findings and implications in the paper do not necessarily reflect those of the sponsors.